\documentclass[runningheads,a4paper]{llncs}
\usepackage{times}
\usepackage{amssymb}
\usepackage{graphicx}
\usepackage{url}
\newcommand{\keywords}[1]{\par\addvspace\baselineskip
\noindent\keywordname\enspace\ignorespaces#1}

\begin{document}

\mainmatter  % start of an individual contribution

\title{On The Differences Between Song and Speech Emotion Recognition: 
Effect of Feature Sets, Feature Types, and Classifiers}

\titlerunning{On The Differences between Song and Speech Emotion Recognition}

\author{Bagus Tris Atmaja\inst{1,2}\and Masato Akagi\inst{2}}
%
% if the names of the authors are too long for the running head, please use the
% format: AuthorA et al.
\authorrunning{Bagus Tris Atmaja and Masato Akagi}

\institute{Sepuluh Nopember Institute of Technology, Indonesia
\and Japan Advanced Institute of Science and Technology, Japan\\ 
\email{bagus@ep.its.ac.id}}


\maketitle

\begin{abstract}
In this paper, we evaluate the different features sets, feature types, and
classifiers on both song and speech emotion recognition. Three feature sets:
GeMAPS, pyAudioAnalysis, and LibROSA; two feature types: low-level descriptors
and high-level statistical functions; and four classifiers: multilayer
perceptron, LSTM, GRU, and convolution neural networks are examined on both song
and speech data with the same parameter values. The results show no remarkable
difference between song and speech data using the same method. In addition,
high-level statistical functions of acoustic features gained higher performance
scores than low-level descriptors in this classification task. This result
strengthen the previous finding on the regression task which reported the
advantage use of high-level features.

\end{abstract}

\section{Introduction}
% contextualizing the problem
Music emotion recognition is an attempt to recognize emotion, either categories
or dimensions, within pieces of music. Music expresses and induces emotion.
Therefore, the universality of emotion within the music can be extracted
regardless the origin of music \cite{fritz2009universal}. Recognizing emotion in
music is important, for instance, in the music application's recommender system.

% what is the problem to be solved
% Are there any existing solution
A Song is part of music that is performed by the human voice. While research 
on music emotion recognition is well-established, research on song 
emotion recognition is less developed. Since it is a part of music, 
recognizing song emotion recognition is essential for music emotion recognition. 
This research aimed to evaluate song emotion recognition in parallel with 
speech emotion recognition.

In speech emotion recognition, several acoustic features, types, and classifier
have been developed. Moore at al. proved that using high-level features improves
the performance of emotion recognition compared to using low-level features
\cite{Moore2014word}. Moreover, Atmaja and Akagi \cite{Atmaja2020a} showed that
using high-level statistical functions (HSF), i.e., Mean+Std, of low-level
descriptors (LLDs) of pyAudioAnalysis feature set \cite{Atmaja2020a} gains
higher performance than using LLD itself. The reported results are obtained in
dimensional emotion recognition tasks. In categorical emotion recognition, no
report showed the effectiveness of Mean+Std from categorical emotion
recognition. In addition to the evaluation of different types of features (LLDs
vs HSFs) in categorical speech emotion recognition, we added the task to
evaluate that feature types evaluation for song emotion recognition, as well as
effect of different feature types and classifiers.

The contribution of this paper, besides the feature types evaluation, is an
evaluation of a feature set derived from LibROSA toolkit \cite{McFee2019}. We
compared both LLDs and HSF from selected LibROSA acoustic features to GeMAPS
\cite{Eyben} and pyAudioAnalysis \cite{Giannakopoulos2015}. We expect an
improvement of performance from those two feature sets by utilizing a larger
number of acoustic features, particularly on HSF feature type. Although the data
is relatively small, i.e., about 1000 utterances for both song and speech, we
choose deep learning-based classifiers to evaluate those features due to its
simplicity. Other machine learning method, e.g., support vector machine, may
obtain higher performance due to its effectiveness on smaller data.

% which is the best
% what is the main limitation
% what do we hope to achieve

% contribution (1) evaluation of hfs on categorical emotional speech and song data
% (2) a proposal of a feature set for speech and emotional song data based on 
% LibROSA toolkit

\section{Dataset}
The Ryerson Audio-Visual Database of Emotional Speech and Song (RAVDESS) 
dataset from Ryerson University is used. This dataset contains multimodal 
recordings of emotional speech and song on both audio and video formats. Speech 
includes seven emotion categories: calm, happy, sad, angry, fearful, surprise, 
and disgust expressions; a neutral with a total of 1440 utterances. 
Song includes five emotion categories: calm, happy, sad, angry, and fearful;
and a neutral with a total of 1012 utterances. Both speech and song are 
recorded at 48 kHz. The detail of the dataset can be found in \cite{Livingstone2018}.

\section{Methods}
Three main methods are evaluated for both emotional speech and song: three 
different feature sets, two feature types for each feature set, and four 
classifiers. The following three sections describe each of those methods.

\subsection{Feature Sets}
% GeMAPS + OpenSMILE
The first evaluated feature set for emotional speech and song is 
Geneva Minimalistic Acoustic Parameter Set (GeMAPS) \cite{Eyben}. This 
feature set is a proposal to standardize acoustic features for voice 
and affective computing. Twenty-tree acoustic features,
known as low-level descriptor (LLD), are chosen 
as minimalistic parameter set while the extended version 
contains added spectral and frequency 
related parameters and its functional. The extended version (eGeMAPS) 
consist of 88 parameters. This research used the minimalistic GeMAPS 
feature set due to its effectiveness compared to other features sets 
\cite{Eyben,Atmaja2020}.

The openSMILE toolkit \cite{Eyben2013} was used to extract 23 LLDs GeMAPS
feature set for each time frame. That frame-based processing is 
conducted with 25 ms window length and 10 ms hop 
length resulting (523, 23) feature size for speech data and (633, 23) feature 
size for song data.

% pyAudioAnalysis
The second evaluated feature set is pyAudioAnalysis (pAA). pyAudioAnalysis 
was designed for general-purpose open-source Python library for audio 
signal analysis. The library provides a wide range of audio analysis procedures
including: feature extraction, classification of audio signals, supervised and
unsupervised segmentation, and content visualization \cite{Giannakopoulos2015}. 
Thirty-four LLDs are extracted on frame-based processing from this feature 
set. Those LLDs, along with the previous GeMAPS and next LibROSA feature sets, 
are shown in Table \ref{tab:lld}. 

\begin{table}[htpb]
\begin{center}
\caption{Acoustic features sets used to evaluate song and speech 
emotion recognition.}
\label{tab:lld}
\begin{tabular}{p{2.5cm} p{7.5cm}}
\hline
Feature set    &   LLDs \\
\hline
GeMAPS & intensity, alpha ratio, Hammarberg index, spectral slope 0-500 Hz, spectral slope 
    500-1500 Hz, spectral flux, 4 MFCCs, F0, jitter, shimmer, Harmonics-to-Noise Ratio (HNR), 
    harmonic difference H1-H2, harmonic difference H1-A3, F1, F1 bandwidth, F1 
    amplitude, F2, F2 amplitude, F3, and F3 amplitude. \\ 
    \hline
pAudioAnalysis & zero crossing rate, energy, entropy of energy, spectral centroid, 
    spectral spread, spectral entropy, spectra flux, spectral roll-off, 
    13 MFCCs, 12 chroma vectors, chroma deviation.\\
    \hline
LibROSA & 40 MFCCs, 12 chroma vectors, 128 mel-scaled spectrograms, 
          7 spectral contrast features, 6 tonal centroid features. \\
\hline
\end{tabular}
\end{center}
\end{table}

% LibROSA
As the final feature set, we selected five LLDs from LibROSA feature 
extractor including MFCCs, chroma, mel spectrogram, spectral 
contrast and tonnetz. The number of features for each LibROSA LLD is shown 
in Table \ref{tab:lld} with a total of 193 features. This number of features 
is chosen based on experiments. The detail of LibROSA version used 
in this experiment can be found in \cite{McFee2019}.

\subsection{Feature Types}
% shows the previous is LLDs
The traditional method to extract acoustic feature from a speech is done 
on a frame-based processing method, i.e., the aforementioned LLDs 
in each acoustic feature set. The higher level acoustic features 
can be extracted as statistical aggregation functions over LLDs on 
fixed-time processing, e.g., an average feature value of each 100 ms, 500 ms, 1 s, or 
per utterance. This high-level statistical functions (HSF) is intended 
to roughly describe the temporal variations and contours of the different LLDs 
over a fixed-time or an utterance \cite{Mirsamadi2017}. In dimensional 
speech emotion recognition, this HSF feature performs better than LLDs, 
as reported in \cite{Atmaja2020a,Schmitt2018,Moore2014word}.

% explain HSF, show supporting papers
Schmitt and Schuller \cite{Schmitt2018} evaluated mean and standard deviation 
(Mean+Std) of GeMAPS feature set and compared it with eGeMAPS and bag-of-audio-word 
(BoAW) representations of LLDs. The result showed that Mean+Std works best 
among the three. Based on this finding, we incorporated Mean+Std as HSFs from 
the previously explained three feature sets. The Mean+Std is calculated 
per utterance on each feature set resulting difference size/dimension for 
each feature set, i.e. 46-dimensional for GeMAPS, 68-dimensional for pyAudioAnalysis, 
and 386-dimensional for selected LibROSA features. Hence, two feature types 
are evaluated; LLD and HSF, from GeMAPS, pyAudioAnalysis and selected 
LibROSA features.

\subsection{Classifiers}
Four classifiers are evaluated: a dense network (or multi-layer perceptron, MLP), 
a long short-term memory (LSTM) network, a gated recurrent unit (GRU) network, 
and a convolution network. The brief explanations of those networks are 
described below.
\begin{enumerate}
\item MLP: Three dense layers are stacked with 256 units and ReLU activation
function for each layer. The last dense layer is flattened, and a dropout rate
with probability 0.4 is added after it. The final layer is a dense layer with
eight units for speech and six units for song with a softmax activation
function.
\item LSTM: Three LSTM layers are stacked with 256 units each and returned all
values. The rest layers are the same as MLP classifier, i.e., a dropout layer
with probability 0.4 and a dense layer with a softmax activation function.
\item GRU: The GRU classifiers similar to LSTM. The LSTM stack is replaced by
GRU stack without changing other parameters.
\item Conv1D: Three 1-dimensional convolution networks are stacked with 256
units and a ReLU activation function for each layer. The filter lengths
(strides) are 4, 8, 12 for first, second, and third convolution layers. The rest
layers are similar to other classifiers.
\end{enumerate}

\section{Results and Discussion}
We divided our results into two parts, analysis of different feature sets and
feature types (i.e., difference features) and analysis of different classifiers.
Both results are presented in terms of accuracy and unweighted average recall
(UAR). Accuracy is widely used to measure the classification error rate in
balanced/near-balanced data. It defines the number of correctly classified
examples divided by the total number of examples, usually presented in \% (here
we used 0-1 scale). UAR is an average recall from all classes, i.e., the number
of correctly classified positive examples divided by the total number of
positive examples in each class. UAR is widely used to justify classification
method from imbalanced data.

\subsection{Effect of different feature types}
Table \ref{tab:result} shows accuracy and UAR of different feature sets and 
feature types. On different feature types, LibROSA-based acoustic features 
perform best on both emotional song and speech data. On different 
feature sets, HSF features perform better than LLD, except on 
pyAudioAnalysis feature set for speech data. Only in that pyAudioAnalysis 
feature set LLDs of pyAudioAnalysis obtained better accuracy and UAR than 
its HSF. In overall evaluation, our proposal on using Mean+Std of LibROSA-based 
acoustic features perform best among six different features. This finding 
suggests that Mean+Std performs well not only on dimensional speech emotion 
recognition (regression task) but also on categorical song and speech 
emotion recognition. 

\begin{table}[htpb]
\begin{center}
\caption{Accuracy and unweighted average recall (UAR) of different feature sets
and feature types on emotional song and speech based on 10-fold validation; both
are in 0-1 scale.}
\label{tab:result}
\begin{tabular}{p{3.2cm} c c c c} %p{1.2cm} p{1cm} p{1.2cm} p{1cm}}
\hline
Feature & \multicolumn{2}{c}{Song}    & \multicolumn{2}{c}{Speech} \\
        & Accuracy   &   UAR     &   Accuracy     &   UAR \\
\hline
GeMAPS    & 0.637    & 0.592    & 0.602    & 0.614 \\
GeMAPS HSF    & 0.753    & 0.762    & 0.662    & 0.653 \\
pyAudioAnalysis    & 0.592    & 0.619    & 0.731    & 0.701 \\
pyAudioAnalysis HSF    & 0.736    & 0.761    & 0.658    & 0.620 \\
LibROSA    & 0.751    & 0.780    & 0.732    & 0.676 \\
LibROSA HSF    & \textbf{0.820}    & \textbf{0.813}    & \textbf{0.774}    & \textbf{0.781} \\
\hline
\end{tabular}
\end{center}
\end{table}

\begin{figure}
\centering
\includegraphics[width=\textwidth]{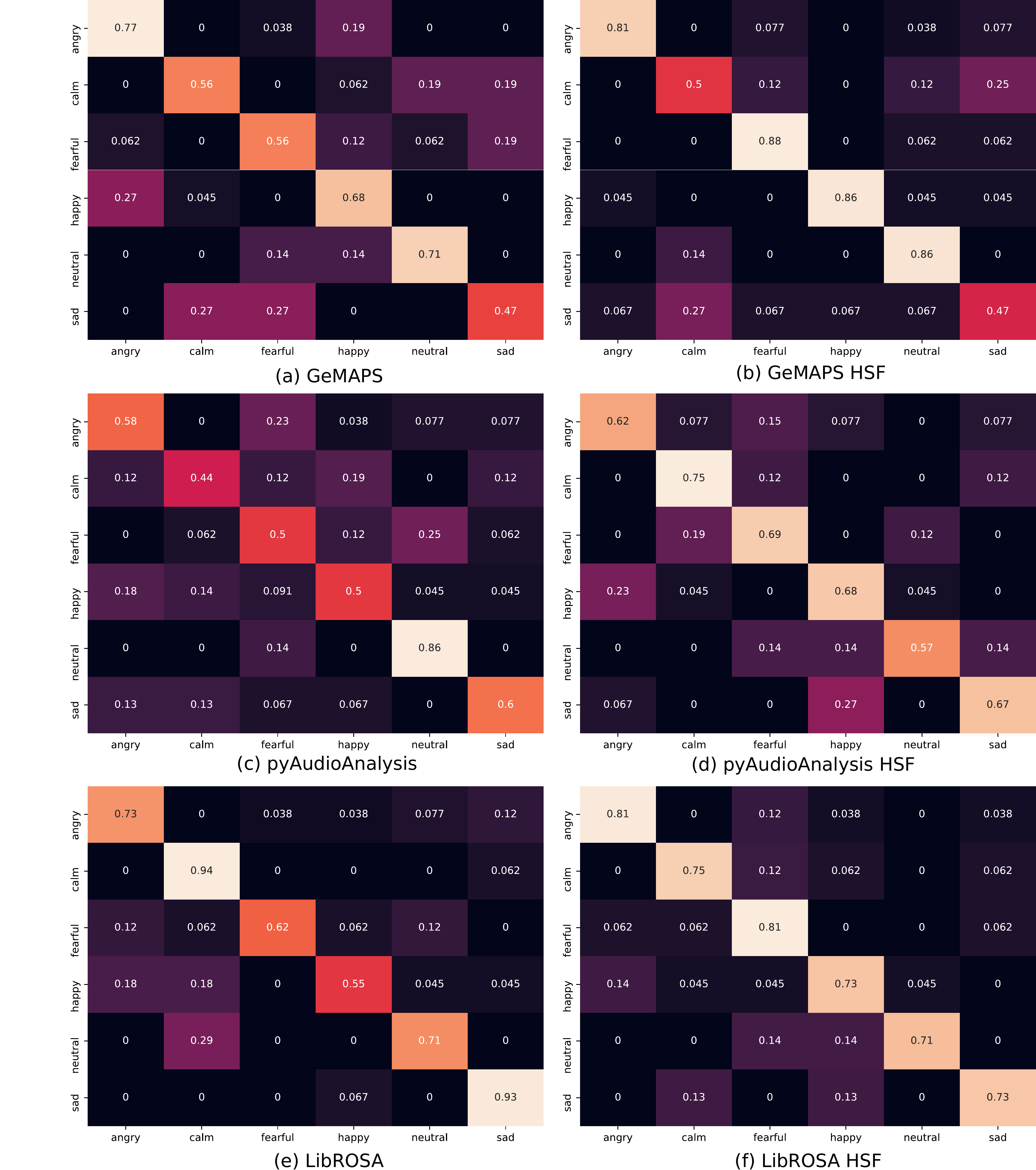}
\caption{Confusion matrix of different evaluated acoustic features on \emph{song} data.}
\label{fig:song_cm}
\end{figure}

\begin{figure}
\centering
\includegraphics[width=\textwidth]{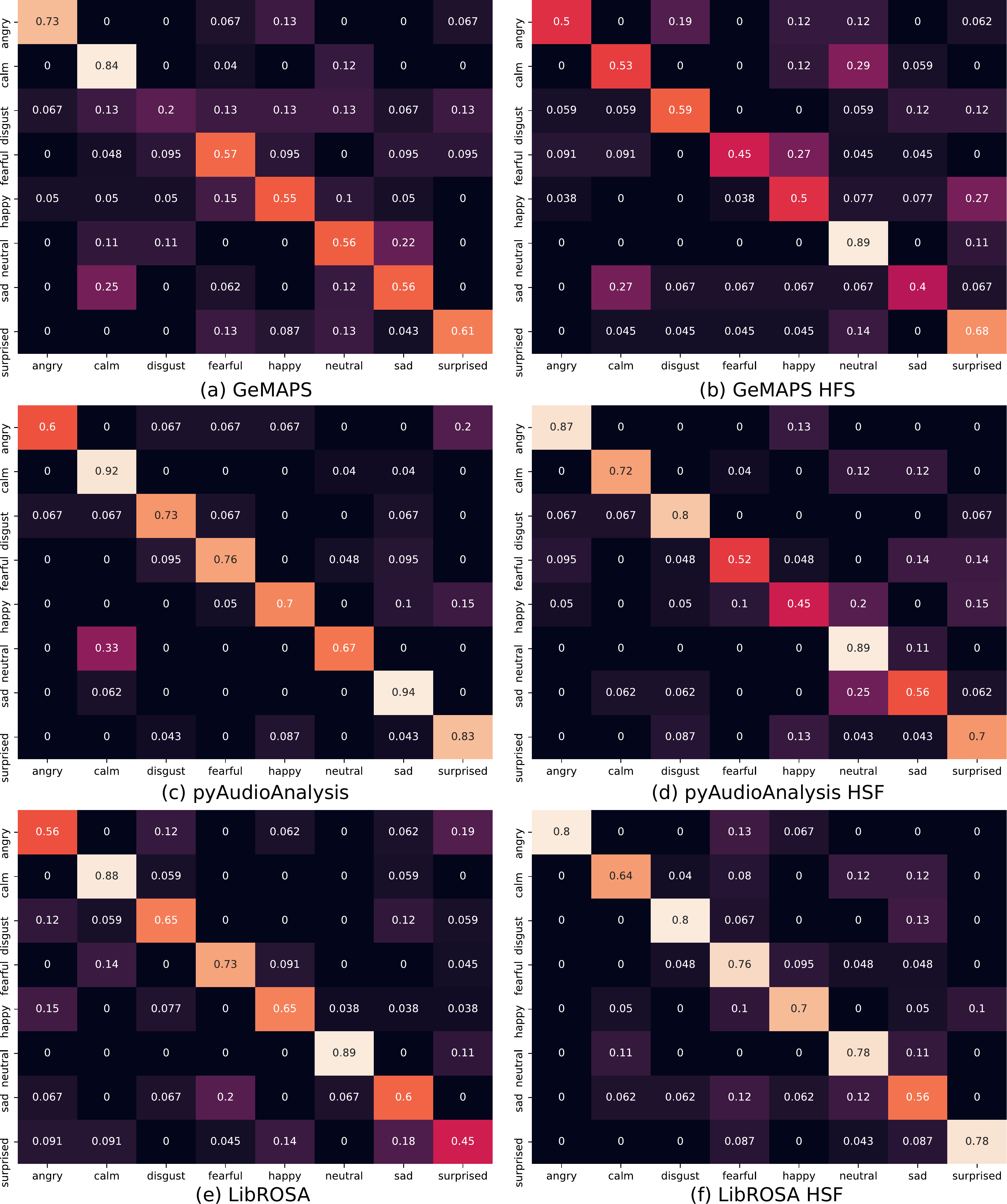}
\caption{Confusion matrix of different evaluated acoustic features on \emph{speech} data.}
\label{fig:speech_cm}
\end{figure}

To find in which emotion category our method performs best and worse, we
performed confusion matrix presentation as shown in Figure \ref{fig:song_cm} and
\ref{fig:speech_cm}. We found feature-specific patterns on the detection of each
emotional song category. For instance, the GeMAPS features, both LLD and HSF,
obtained the lowest recall on the sad category while LibROSA HSF obtained high
recall on this sad category. The happy category which obtained low scores on
LLDs of pyAudioAnalysis and LibROSA obtained improvement on its HSFs. The
highest and lowest recall scores vary for each feature on both song and speech
data. On the speech data (Figure \ref{fig:speech_cm}), LLD features shows best
recall scores on calm category while HSF features obtained better score on
neutral category. On both song and speech data, it was found the HSF improved
recall scores of LLDs features. Since each feature has a different highest
recall score, it is interesting to combine those features sets to improve the
current result for future research direction.

Although we found no remarkable different on the performance trends 
from the same features sets between emotional song and speech data, a significant 
difference maybe found on the using of specific acoustic features, e.g., 
F0 contour, spectral features and amplitude envelope, as reported in \cite{Nguyen2018} 
for emotional singing voice.

\subsection{Effect of different classifiers}
For different classifiers, the accuracy and recall scores are presented in 
Table \ref{tab:classifiers}. On the previous table, the results was obtained 
using LSTM classifiers. In overall evaluation, the LSTM classifier obtained 
the best result among other three. However, GRU classifier shows better 
UAR on song data, i.e., on recognizing each emotion category. This result 
shows that recurrent-based classifiers (LSTM and GRU) performs better 
than MLP and 1-dimensional convolution network on classification 
of emotional song and speech. Similar to the previous Table \ref{tab:result}, 
the song data showed higher scores than speech data. Since the song data 
contains fewer data (samples) and fewer emotion categories than speech data, 
it can be concluded that song is more emotional than speech. The result on 
the same classifier, same feature set, and same feature type supports 
this finding.
\begin{table}
\label{tab:classifiers}
\begin{center}
\caption{Accuracy and unweighted average recall (UAR) of emotional song and
speech on different classifiers using LibROSA HSF feature based on 10-fold
validation; both are in 0-1 scale.}
\begin{tabular}{p{3cm} c c c c} %p{1.2cm} p{1cm} p{1.2cm} p{1cm}}
\hline
Classifier & \multicolumn{2}{c}{Song}    & \multicolumn{2}{c}{Speech} \\
        & Accuracy   &   UAR     &   Accuracy     &   UAR \\
\hline
MLP    & 0.794    & 0.804    & 0.729    & 0.755 \\
LSTM    & \textbf{0.820}    & 0.813    & \textbf{0.785}    & \textbf{0.781} \\
GRU    & 0.812    & \textbf{0.844}    & 0.785    & 0.764 \\
Conv1D    & 0.743    & 0.806    & 0.687    & 0.690 \\
\hline
\end{tabular}
\end{center}
\end{table}

\section{Conclusions}
% first, there is similarity between speech and song
% second, song is more emotional than speech
We presented an evaluation of different feature sets, feature types, and
classifiers for both song and speech emotion recognition. First, we conclude
that there is no remarkable difference between song and speech emotion
recognition on the same features and classifiers based on the evaluated methods.
In other words, the features types/sets and classifiers which gain better
performance on song data will also gain better performance on speech data.
Second, song is more emotional than speech. On both accuracy and unweighted
average recall, the scores obtained by song data always higher than speech data.
Both song and speech data contain the same statements; hence, the different is
the intonation/prosody and other acoustic information which is captured by
acoustic features. Third, on different feature types, high-level statistical
functions consistently performed better than low-level descriptors. For the
future research direction, we planned to combine different acoustic features
types and sets since the result showed differences among emotion categories.

\bibliographystyle{splncs03}
\bibliography{cmmr2020}

\end{document}